\documentclass[prl,aps,twocolumn,showpacs,10pt,a4paper]{revtex4}

\usepackage[dvips]{graphics}
\usepackage[dvips]{graphicx}
\usepackage{float}
\usepackage{epsfig}
\usepackage{tabularx,longtable}
\usepackage{color}
\usepackage{fancybox}
\usepackage{amsmath}
\usepackage{ulem}
\usepackage{pdfpages}

\newcommand{\bi}{\begin{itemize}}
\newcommand{\ei}{\end{itemize}}
\newcommand{\be}{\begin{equation}}
\newcommand{\ee}{\end{equation}}

\newcommand{\bea}{\begin{eqnarray}}
\newcommand{\eea}{\end{eqnarray}}
\newcommand{\beastar}{\begin{eqnarray*}}
\newcommand{\eeastar}{\end{eqnarray*}}

\newcommand{\eq}[1]{Eq.~(\ref{#1})}

\begin{document}

\title{Recovery of free energy branches in single molecule experiments}

\author{Ivan Junier$^1$} 
\author{Alessandro Mossa$^2$}
\author{Maria Manosas$^3$} 
\author{Felix Ritort$^{2,4}$\footnote{To whom correspondence should be addressed. \\  E-mail: ritort@ffn.ub.es.}}
\affiliation{$^1$Programme d'\'Epig\'enomique, 523 Terrasses de l'Agora, 91034 \'Evry, France \\
$^2$Departament de F\'{i}sica Fonamental, Facultat de F\'{\i}sica, Universitat de Barcelona, Diagonal 647, 08028 Barcelona, Spain \\
$^3$Laboratoire de Physique Statistique, \'Ecole Normale Sup\'erieure, 24 rue Lhomond, 75231 Paris, France \\
$^4$ CIBER-BBN, Networking Centre on Bioengineering, Biomaterials and Nanomedicine
}

\begin{abstract}
We present a method for determining the free energy of coexisting states
from irreversible work measurements. Our approach is
  based on a fluctuation relation that is valid for
dissipative transformations in partially equilibrated
systems. To illustrate the validity and usefulness of
  the approach, we use optical tweezers to determine the free
energy branches of the native and unfolded states of a two-state
molecule as a function of the pulling control parameter. We determine, within
$0.6\ k_{\rm B}T$ accuracy, the transition point where the free
  energies of the native and the unfolded states are equal.
\end{abstract}


\maketitle


Recent developments in statistical physics \cite{Jarzynski:1997yh}
have provided new methods to extract equilibrium free energy
differences in small systems from measurements of the mechanical work
in irreversible processes (see \cite{kurchan:2007uh,ritort:2008ik} for
reviews). In this regard, fluctuation relations \cite{kurchan:2007uh}
are generic identities that establish symmetry properties for the
probability of exchanging a given amount of energy between the system
and its environment along irreversible processes.  If a system,
initially in thermodynamic equilibrium, is strongly perturbed by fast
varying a control parameter $\lambda$ between two values $\lambda_0$
and $\lambda_1$, then the system is driven out of
equilibrium. The work exerted upon the system, averaged over the ensemble of
all possible trajectories, reads $\langle W \rangle= \langle
\int_{\lambda_0}^{\lambda_1} \; (\partial{\cal H}/\partial\lambda) d\lambda
\rangle$ where ${\cal H}$ is the system Hamiltonian. According to the second law of thermodynamics, $\langle W
\rangle$ is always greater than the free energy difference between the
initial and final states, $\Delta G=G(\lambda_1)-G(\lambda_0)$. The
Crooks fluctuation relation \cite{Crooks:2000ez} extends the
predictive power of the Second Law by establishing a symmetry relation
for arbitrary functionals of a trajectory $\Gamma$ measured along a
nonequilibrium process (forward or F process) and its time reversed
one (reverse or R process). In the forward process the system starts
in equilibrium at $\lambda_0$ and $\lambda$ is varied from $\lambda_0$
to $\lambda_1$ for a time $t_f$ according to an arbitrary protocol
$\lambda(t)$ (i.e., $\lambda_1=\lambda(t_f)$). In the reverse process
the system starts in equilibrium at $\lambda_1$ and $\lambda$ is
varied from $\lambda_1$ to $\lambda_0$ following the time reversed
scheme, given by $\lambda(t_f-t)$. In its most general form the Crooks
fluctuation relation reads \cite{Crooks:2000ez} \be \langle
\mathcal{F}\exp[-\beta(W-\Delta G)] \rangle_\mathrm{F} = \langle
\hat{\mathcal{F}} \rangle_\mathrm{R} \,,
\label{cro}
\ee 
where $\mathcal{F}$ stands for an arbitrary functional of the
forward trajectories the system can take through phase space, $\beta$ is
the inverse of the thermal energy $k_\mathrm{B}T$  where
$k_\mathrm{B}$ is the Boltzmann constant and $T$ the temperature of
  the environment. In this relation, $\hat{\mathcal{F}}$ is the time
reversal of $\mathcal{F}$, while the averages
$\langle\bullet\rangle_{\mathrm{F}(\mathrm{R})}$ are taken over the
ensemble of all possible forward (reverse) trajectories. The
particular case $\mathcal{F}=\delta(W-W(\Gamma))$ yields a relation
between work distributions along the forward and reverse processes,
$P_\mathrm{F}(W)=P_\mathrm{R}(-W)\exp[\beta(W-\Delta G)]$. This relation
has been experimentally tested and used to extract free energy
differences in single molecule experiments
\cite{Hummer:2001,Collin:2005rb,Imparato:2008}. A thorough discussion on its validity domain can be found in \cite{Crooks:2000ez}

\paragraph{ Fluctuation relation under partial
equilibrium conditions.}  By considering only the trajectories that go
from one specific subset of configurations to another one, Maragakis
et al.\ \cite{Maragakis:2008hj} have derived another relation useful
to extract free energy differences between subsets of states.  In
principle, the validity of \eq{cro} is restricted to initial
conditions that are Gibbsian over the whole phase space $\mathcal{S}$
(what we might call {\it global} thermodynamic equilibrium). It is,
however, possible to extend \eq{cro} to the case where the initial
state is Gibbsian but restricted over a subset of configurations (what
we might call {\it partial} thermodynamic equilibrium). A
relation mathematically similar to \eq{cro} can be derived, but involving
nonequilibrium processes that are in partial (rather than global)
equilibrium. It is useful to rephrase here the
derivation in such a way to emphasize the role played by
partial equilibrium.  As we will see this makes it possible to
experimentally determine the free energy of coexisting states for
values of $\lambda$ such that the system is never globally
  equilibrated.

Let $P_{\lambda,\mathcal{S'}}^{\rm eq}({\cal C})$ denote the partially
equilibrated (i.e., Boltzmann--Gibbs) distribution for a given value of
$\lambda$. Such distribution is restricted over a subset
$\mathcal{S'}$ of configurations ${\cal C}$ contained in
$\mathcal{S}$ (i.e., ${\cal C}\in \mathcal{S'}\subseteq \mathcal{S}$). The case
$\mathcal{S'}=\mathcal{S}$ corresponds to global equilibrium:
$P_{\lambda}^{\rm EQ}({\cal C})\equiv P^{\rm eq}_{\lambda,\mathcal{S}}({\cal C})$. Partially equilibrated states satisfy
$P^{\rm eq}_{\lambda,\mathcal{S'}}({\cal C})=P^{\rm EQ}_{\lambda}({\cal
  C})\chi_{\mathcal{S'}}({\cal C}){\cal Z}_{\lambda,\mathcal{S}}/{\cal
  Z}_{\lambda,\mathcal{S'}}$, where $\chi_{\mathcal{S'}}$ is the
characteristic function defined over the subset $\mathcal{S'}$
  ($\chi_{\mathcal{S'}}({\cal C})=1$ if ${\cal C}\in \mathcal{S'}$ and
  zero otherwise), and ${\cal  Z}_{\lambda,\mathcal{S'}}$ is
the partition function restricted to the subset $\mathcal{S'}$, i.e.,  
${\cal  Z}_{\lambda,\mathcal{S'}}=\sum_{{\cal
    C}\in\mathcal{S'}}\exp(-\beta E_{\lambda}({\cal C}))$, with
$E_{\lambda}({\cal C)}$ the energy function of the system for a given
  $\lambda$ and ${\cal C}$. Given a forward trajectory $\Gamma$, going from
  configuration ${\cal C}_0$ when $\lambda=\lambda_0$ to ${\cal C}_1$ for $\lambda=
  \lambda_1$, let ${\cal S}_0$ (${\cal S}_1$) be the subset of ${\cal S}$ over which 
  the system is partially equilibrated at $\lambda_0$ ($\lambda_1$). 
  Consider now the following transformation of the functional $\mathcal{F}$ in \eq{cro}:
$\mathcal{F}(\Gamma)\to
\chi_{\mathcal{S}_0}(\mathcal{C}_0)\mathcal{F}(\Gamma)\chi_{\mathcal{S}_1}(\mathcal{C}_1)$. 
Under previous conditions the following identity can be proved (see Supp. Mat.):
\be
\frac{ \langle \mathcal{F}\exp[-\beta(W-\Delta G_{\mathcal{S}_0,\lambda_0}^{\mathcal{S}_1,\lambda_1})] \rangle_\mathrm{F}^{\mathcal{S}_0\to\mathcal{S}_1} } 
{ \langle \hat{\mathcal{F}}  \rangle_\mathrm{R}^{\mathcal{S}_0\gets\mathcal{S}_1} }
=
\frac{ p_\mathrm{R}^{{\cal S}_0\gets {\cal S}_1}}
{ p_\mathrm{F}^{{\cal S}_0\to {\cal S}_1} } 
\,,
\label{cro2}
\ee
where the average $\langle\bullet\rangle^{\mathcal{S}_0\to\mathcal{S}_1(\mathcal{S}_0\gets\mathcal{S}_1)}_{\mathrm{F}(\mathrm{R})}$ is now restricted to forward (reverse) trajectories that
start in partially equilibrated state $\mathcal{S}_0$ ($\mathcal{S}_1$)
at $\lambda_0$ ($\lambda_1$) and end in $\mathcal{S}_1$ ($\mathcal{S}_0$)
at $\lambda_1$ ($\lambda_0$). 
$p_\mathrm{F}^{{\cal S}_0\to{\cal S}_1}$ ($p_\mathrm{R}^{S_0\gets S_1}$) stands for the probability to be in ${\cal S}_1$ (${\cal S}_0$) at the end of the
forward (reverse) process defined above, and $\Delta G_{\mathcal{S}_0,\lambda_0}^{\mathcal{S}_1,\lambda_1}=G_{\mathcal{S}_1}(\lambda_1)-G_{\mathcal{S}_0}(\lambda_0)$
is the free energy difference between partially equilibrated states
$\mathcal{S}_0$ and $\mathcal{S}_1$.  In the following, we will 
drop the subscript (F, R), leaving the direction of the arrow to distinguish forward from reverse. Moreover, we will adopt the shorthand notation $\mathcal{P}_{{\cal S}_0}^{{\cal S}_1}\equiv p^{{\cal S}_0\to{\cal S}_1}/p^{{\cal S}_0\gets {\cal S}_1}$. 
 If $\mathcal{F}=1$ we obtain a generalization of the Jarzynski equality, 
 $\mathcal{P}_{{\cal S}_0}^{{\cal S}_1}\langle \exp[-\beta(W-\Delta G_{\mathcal{S}_0,\lambda_0}^{\mathcal{S}_1,\lambda_1})]\rangle=1$. Whereas for the particular case
  $\mathcal{F}=\delta(W-W(\Gamma))$, we get the relation
\be
\mathcal{P}_{{\cal S}_0}^{{\cal S}_1}\frac{P^{{\cal S}_0\to {\cal S}_1}(W)}{P^{{\cal S}_0\gets {\cal S}_1}(-W)}=
\exp[\beta(W-\Delta G_{\mathcal{S}_0,\lambda_0}^{\mathcal{S}_1,\lambda_1})] \,,
\label{jun}
\ee
which has been used in \cite{Maragakis:2008hj} in the case of global equilibrium initial conditions.

\begin{figure}[t]
\includegraphics[scale=0.41]{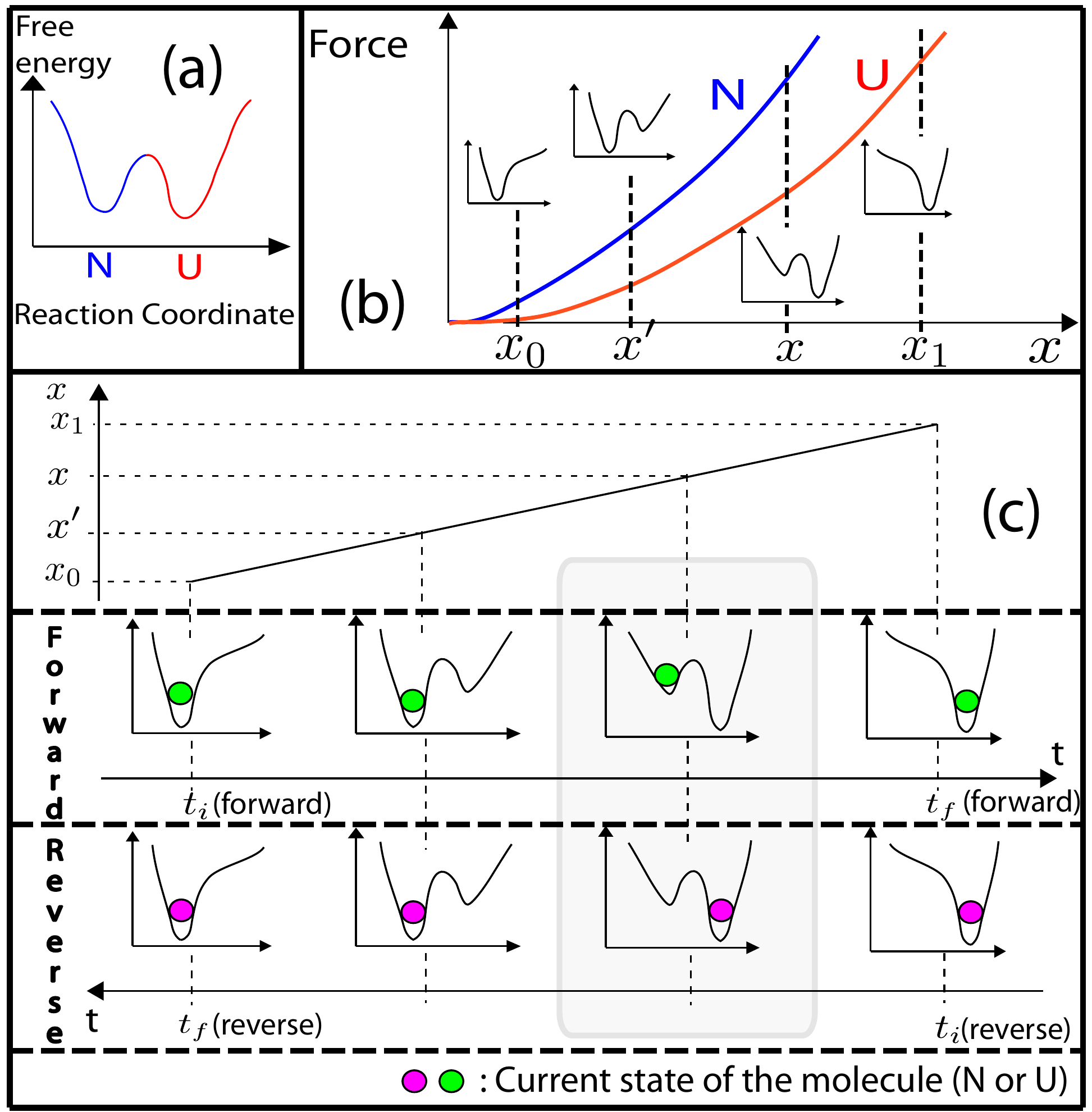}
\caption{{\it A two-state molecule in a pulling experiment}. (a)
  Schematic picture of the free energy landscape in a two-state folder
  as a function of the reaction coordinate. The blue and red colors
  represent the subsets of configurations that define the $N$ and $U$
  states respectively. (b) Depending on the value of the control
  parameter $x$, the shape of the free energy landscape is tilted
  toward one state (either $N$ or $U$). For each value of $x$, the
  Boltzmann--Gibbs equilibrium value of the force restricted to each
  state defines the native (blue) and unfolded (red) force-distance
  branches. (c) During a ramping protocol, $x$ is changed at a constant
  pulling speed from $x_0$ to $x_1$ (from $x_1$ to $x_0$ in the reverse
  case) and the molecule can visit both states as indicated by the
  colored circles. In order to measure $G_U(x)$ (all free energies are computed
  with respect to $G_N(x_0)$)
  at a given value of $x$ (gray area), only forward and reverse
  trajectories that are in $N$ at $x_0$ and in $U$ at $x$ have to be
  considered (which is true only for the reverse trajectory in the
  example shown in the picture).}
\label{fig1}
\end{figure}

\begin{figure}
\includegraphics[width=8cm]{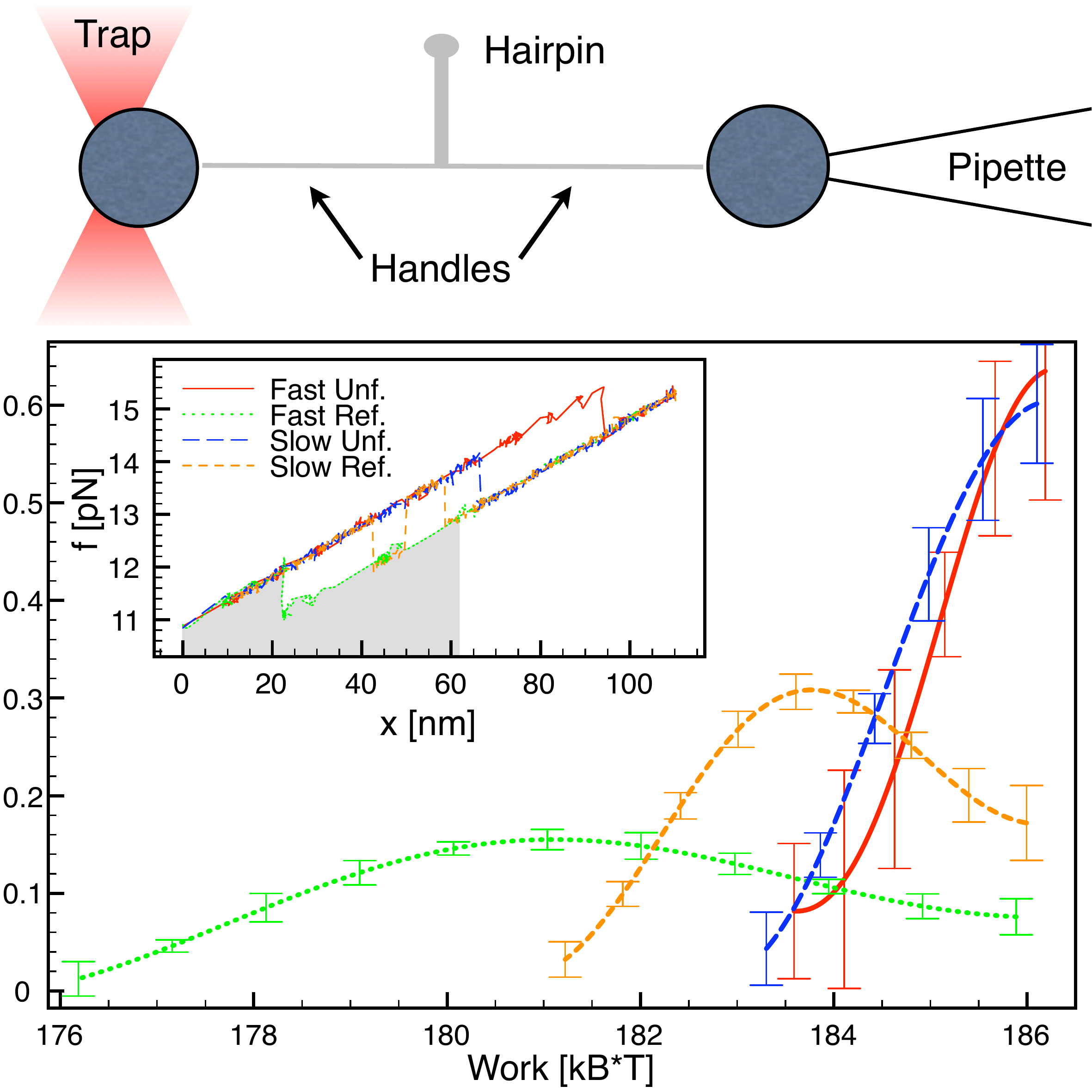}
\caption{(Upper panel) Experimental setup. (Lower panel) Unfolding and
  refolding work
  distributions $P^{N\to U}(W)$, $P^{N\gets U}(-W)$ at
  $x=\bar{x}=61.75$ nm measured at
  two pulling speeds: 300 nm/s (red -unfolding- and green
  -refolding-) and 40 nm/s (blue -unfolding-
  and orange -refolding-). In the inset we show two force-distance 
  cycles corresponding to
  low (40 nm/s) and fast (300 nm/s) pulling speeds. The gray area 
   indicates the mechanical work ($=\int_{\bar x}^{x_0} f dx$) exerted along the green refolding trajectory.
  Statistics (number of
  molecules, total number of unfolding/refolding cycles): 40 nm/s
  (2, 223), 50 nm/s (2, 183), 300 nm/s (3, 337), 400 nm/s (1, 551). More details about the data analysis procedure can be found in the Supp. Mat.}
\label{fig2}
\end{figure}

\paragraph{Experimental test.}
Here we test the validity of \eq{jun} by performing
single molecule experiments using optical tweezers. Let us consider an
experiment where force is applied to the ends of a DNA
hairpin that unfolds/refolds in a two-state manner.  The conformation
of the hairpin can be characterized by two states, the unfolded state
($U$) and the native or folded state ($N$) --- see
Fig.~\ref{fig1}. The thermodynamic state of the molecule can be
controlled by moving the position of the optical trap relative to a
pipette (Fig.~\ref{fig2}, upper panel). The relative position of the
trap along the $x$-axis defines the control parameter in our
experiments, $\lambda= x$.  Depending on the value of $x$ the molecule
switches between the two states according to a rate that is a function
of the instantaneous force applied to the molecule
\cite{Evans:1997np}. In a ramping protocol the value of $x$ is changed
at constant pulling speed from an initial value $x_0$ (where the
molecule is always folded) to a final value $x_1$ (where the molecule
is always unfolded) and the force $f$ (measured by the optical trap) versus distance $x$ curves
recorded.  By computing i) the fraction of forward
trajectories (i.e., increasing $x$) that go from $N$ ($\equiv {\cal
  S}_0$) at $x_0$ to $U$ ($\equiv {\cal S}_1$) at $x$ and ii) the
fraction of reverse trajectories (decreasing extension) that go from
$U$ at $x$ to $N$ at $x_0$, we can determine
$\mathcal{P}_{N}^{U}$. Then by measuring the corresponding work values
for each of these trajectories, we can use \eq{jun} to
estimate the free energy of the unfolded branch $G_U(x)$ as a function
of $x$. By repeating the same operation with $N$ instead of $U$, the
free energy of the folded branch $G_N(x)$ can be measured as
well. Note that we adopt the convention of measuring all free energies
with respect to the free energy $G_N(x_0)$ of the native state at
$x_0$. We are also able to compute the free energy difference between
the two branches, $\Delta G_N^U(x)=G_U(x)-G_N(x)$.

We have pulled a 20 bps DNA hairpin using a miniaturized dual-beam
laser optical tweezers apparatus \cite{hairpin}.  Molecules have been
pulled at two low pulling speeds (40 and 50 nm/s) and two fast pulling
speeds (300 and 400 nm/s), corresponding to average loading rates
ranging between 2.6 and 26 pN/s, from $x_0=0$ to $x_1=110.26$ nm (for
convenience we take the initial value of the relative distance
trap-pipette equal to 0). A few representative force-distance curves
are shown in Fig.~\ref{fig2} (inset of lower panel). We have then
selected a value $x=\bar{x}=61.75$ nm, close to the expected
coexistence value of $x$ where both $N$ and $U$ states have the same
free energy (see below). Such value of $x$ is chosen in order to have good
statistics for the evaluation of the unfolding and refolding work distributions. The
system is out of equilibrium at the four pulling speeds. To extract
the free energy of the unfolded branch $G_U(\bar{x})$, we have
measured the work values $W^{N\to U}=\int_{x_0}^{\bar{x}}f dx$,
$W^{N\gets U}=\int_{\bar{x}}^{x_0}f dx$ along the unfolding and
refolding trajectories, respectively, and then determined the
distributions $P^{N\to U}(W)$ and $P^{N\gets U}(-W)$. In the main
panel of Fig.~\ref{fig2} we show the work distributions obtained for a
slow and fast pulling process. Note that the support of the unfolding
work distributions is bounded by the maximum amount of work that can
be exerted on a molecule between $x_0$ and $\bar{x}$. This bound
corresponds to the work of those unfolding trajectories that have
never unfolded before reaching $\bar{x}$.

\begin{figure}[t]
\includegraphics[width=8cm]{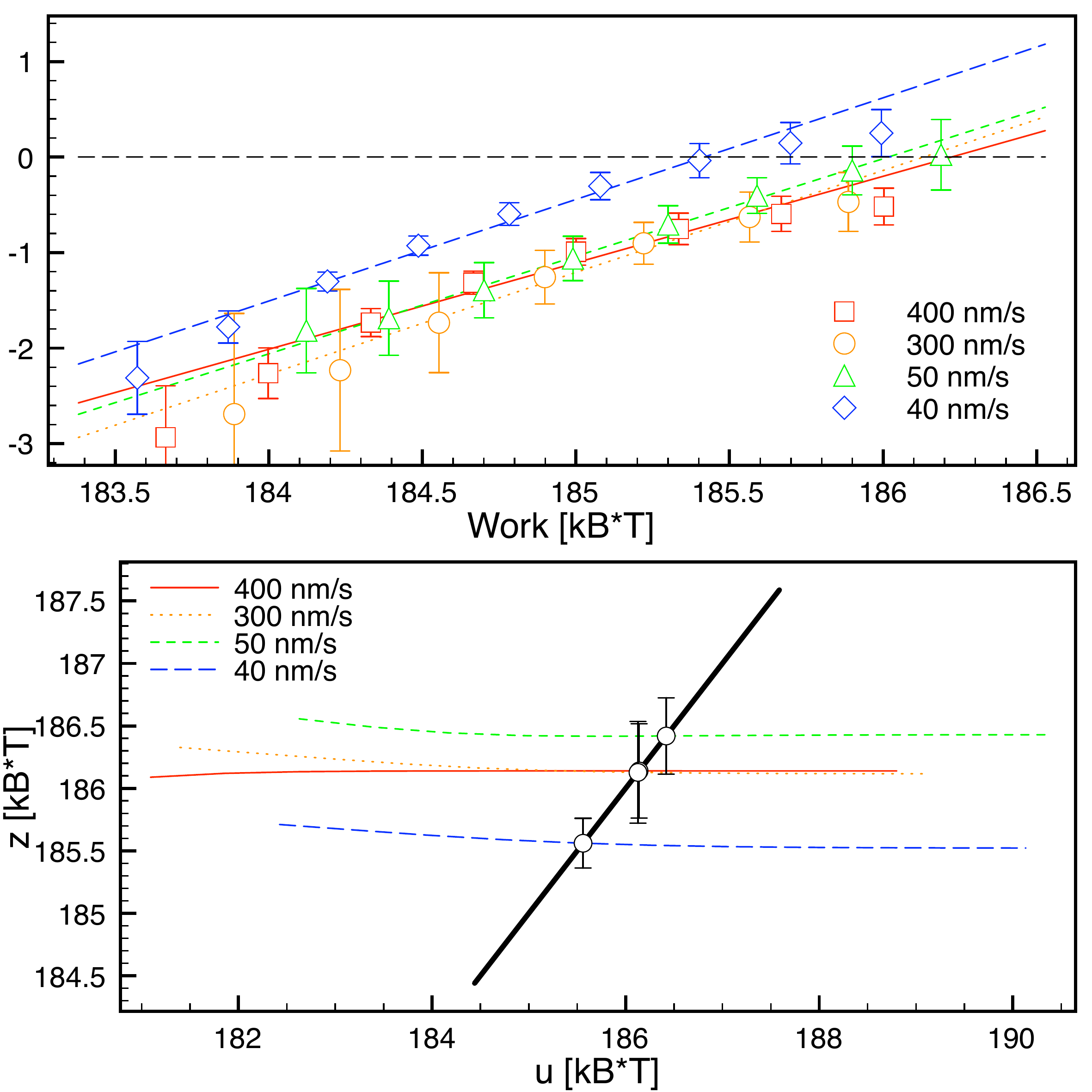}
\caption{{\it Experimental verification of the fluctuation relation
  \eq{jun}}. (Upper panel) Plot of $\log\mathcal{P}^{U}_{N}(\bar{x})+
  \log(P^{N\to U}(W)/P^{N\gets U}(-W))$ as
  a function of $W$ (in $k_\mathrm{B}T$ units) at different pulling speeds. A
  least squares fitting method of the experimental data to straight lines gives the following
  slopes: $0.91(4)$ (400 nm/s), $1.07(8)$ (300 nm/s), $1.02(6)$
  (50 nm/s), $1.07(3)$ (40 nm/s). (Lower panel) Bennett's acceptance
  ratio method. We plot the function $z(u)$ \eq{Bennett} for different
  pulling speeds. The thick black line corresponds to the function $z=u$. Error bars correspond to the standard deviation of the
Bennett estimate, as determined in Ref.~\cite{Shirts:2003ok}. Data statistics are reported in the
  caption of Fig.~\ref{fig2}.}    
\label{fig3}
\end{figure}

As a direct test of the validity of \eq{jun}, in the upper panel of
Fig.~\ref{fig3} we plot the
quantity $\log(P^{N\to U}(W)/P^{N\gets
  U}(-W))+\log\mathcal{P}^{U}_{N}(\bar{x})$ against $W$ (in $k_\mathrm{B}T$ units). As expected, all data fall into straight lines of slope
close to 1. The intersections of such lines with the $W$-axis provide
an estimate of $G_U(\bar{x})$. Note that both fast and slow pulling
speeds intercept the horizontal axis around the same value within 0.75
$k_\mathrm{B}T$ of error. With this method, we estimate
$G_U(\bar{x})=185.9(5)\ k_\mathrm{B}T$. Note that this free energy estimate, as all the others in this paper, refers to the whole system, comprising hairpin, handles, and trap.

A more accurate test of the validity of \eq{jun} and a better estimation \cite{Shirts:2003ok}
of the free energy $G_U(\bar{x})$  can be obtained through the Bennett acceptance ratio method
\cite{Bennett:1976ju}. In Bennett's method we define the following
functions:
\beastar
z_\mathrm{UNF}(u)&=&\log\left\langle\frac{1}{1+\frac{n_\mathrm{UNF}}{n_\mathrm{REF}}\exp[\beta(W-u)]}\right\rangle_\mathrm{UNF} \,, \\
z_\mathrm{REF}(u)&=&\log\left\langle\frac{\exp(-W)}{1+\frac{n_\mathrm{UNF}}{n_\mathrm{REF}}\exp[-\beta(W+u)]}\right\rangle_\mathrm{REF} \,.
\eeastar
From \eq{jun} it has been proved \cite{Bennett:1976ju,Shirts:2003ok} 
that the solution of the equation $z(u)=u$, where
\be \label{Bennett}
z(u)\equiv z_\mathrm{REF}(u)-z_\mathrm{UNF}(u)-\log\mathcal{P}^{U}_{N}(\bar{x}) \,,
\ee
is the optimal (minimal variance) estimate of $G_U(\bar{x})$.

\begin{figure}[t]
\includegraphics[width=8cm]{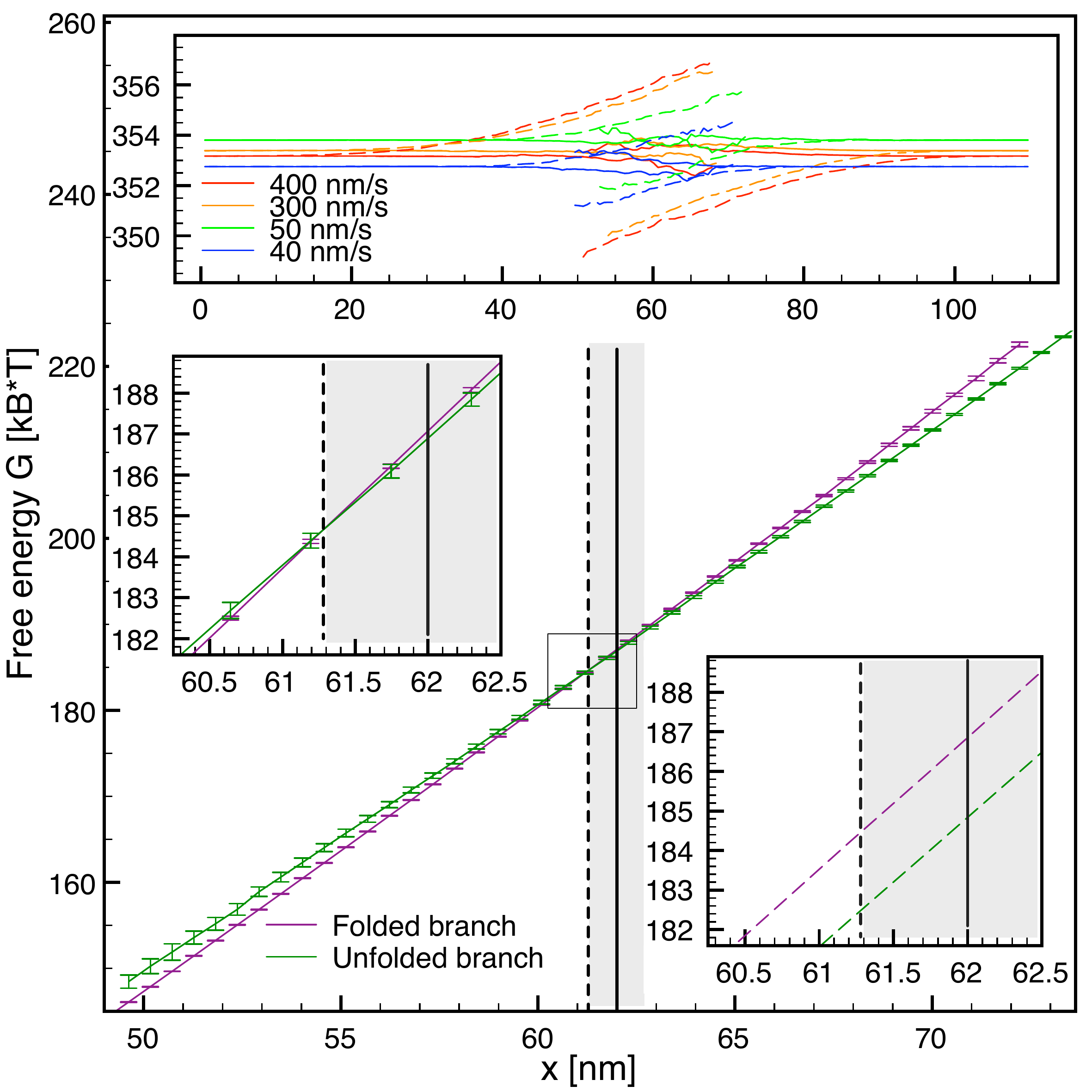}
\caption{{\it Free energy of the folded and unfolded branches}. The purple line is $G_N(x)$, the dark green line is $G_U(x)$, both computed with respect to $G_N(x_0)$.  The black solid line and the gray area mark the expected value $x_\mathrm{c} = 62.0(7)$ nm and its standard error, while the
 dashed black line is the crossing point $x_\mathrm{c} \approx 61.28$ nm. Both left and right insets show a magnified image of the coexistence region. In the right one, the dashed lines represent the free energies that we would obtain if we had overlooked the factor $\mathcal{P}^{U}_{N}$. The top panel shows the functions $\Delta G^U_N(x)-\Delta G^U_U(x)$ and $\Delta G^N_N(x)-\Delta G^N_U(x)$, which, by definition, should be independent of $x$ and equal to $G_U(x_1)$. Again, the dashed lines represent the result if we do not include the factor $\mathcal{P}^{U}_{N}$.}
\label{fig4}
\end{figure}

In Fig.~\ref{fig3} (lower panel) we plot the function $z(u)$ for
different pulling speeds. It is quite clear that the functions $z(u)$ are 
approximately constant along the $u$ axis
and cross the line $z=u$ around the same value $G_U(\bar{x})=186.0(3)$. 
A distinctive aspect of \eq{jun} is the presence of the
factor $\mathcal{P}^{U}_{N}$. If such correction was not taken into account
  then the fluctuation relation would not be satisfied anymore. We have
  verified that if ${P}^{U}_{N}$ is not included in the analysis,
  then the Bennett acceptance ratio method gives free energy estimates
  that depend on the pulling speed (see Supp. Mat.).

\paragraph{Free energy branches.} After having verified that the fluctuation relation \eq{jun} holds and
that it can be used to extract the free energy of the unfolded
($G_U(\bar{x})$) and folded ($G_N(\bar{x})$, data not shown)
branches, we have repeated the same procedure in a wide range of $x$
values. The range of values of $x$ is such that at least 8
trajectories go through $N$ (or $U$) (ensuring that we get a
reasonable statistical significance). The results of the
reconstruction of the free energy branches are shown in
Fig.~\ref{fig4}.  The two free energy branches cross each other at a
coexistence value $x_c$ at which the two states ($N$ and $U$) are equally
probable. This is defined by $\Delta G_N^U(x_c)=0$. We get $x_\mathrm{c}\approx
61.28$ nm, which is in good agreement with another estimate, 
$x_\mathrm{c} = 62.0(7)$ nm, obtained interpreting 
experimental data according to a simple phenomenological model (see Supp. Mat.).  In the insets of Fig.~\ref{fig4} we zoom the
crossing region. It is interesting to recall again the importance of
the aforementioned correction term ($\mathcal{P}^{U}_{N}$) to \eq{jun}. If such term is not
included in the analysis then the two reconstructed branches never
cross (bottom right inset). This result is incompatible with the
existence of the unfolding/refolding transition in the hairpin,
showing that the factor $\mathcal{P}^{U}_{N}$ is key to measure free
energy branches.

Equation (\ref{cro2}) is valid in the very general situation of partially
equilibrated initial states which, however, are arbitrarily far from
global equilibrium. This makes the particular case \eq{jun} a very
useful identity to recover the free energy of states that cannot be
observed in conditions of thermodynamic global equilibrium. We have
shown how it is possible to apply \eq{jun} to recover free energy
differences of thermodynamic branches of folded and unfolded states in
a two-state DNA hairpin.  These methods can be further extended to the
recovery of free energies of non-native states such as
misfolded or intermediates states. 

\begin{acknowledgments}
We are grateful to M. Palassini for a careful reading of the manuscript. We acknowledge financial support from grants FIS2007-61433, NAN2004-9348, SGR05-00688 (A.M, F.R). 
\end{acknowledgments}

\bibliographystyle{phaip}

\begin{thebibliography}{10}

\bibitem{Jarzynski:1997yh}
C.~{Jarzynski},
\newblock \prl {\bf 78}, 2690 (1997).

\bibitem{ritort:2008ik}
F.~Ritort,
\newblock Adv.\ Chem.\ Phys.\ {\bf 137}, 31 (2008).

\bibitem{kurchan:2007uh}
J.~{Kurchan},
\newblock J.\ Stat.\ Mech.\ (2007) P07005.

\bibitem{Crooks:2000ez}
G.~E.~{Crooks},
\newblock \pre {\bf 61}, 2361 (2000).

\bibitem{Hummer:2001} G. Hummer and A. Szabo, Proc. Nat. Acad. Sci (USA)
  {\bf 98}, 3658 (2001).

\bibitem{Collin:2005rb}
D.~Collin et~al.,
\newblock Nature (London) {\bf 437}, 231 (2005).

\bibitem{Imparato:2008} A. Imparato, F. Sbrana and M. Vassalli, Europhys. Lett. {\bf 82} 58006 (2008). 


\bibitem{Maragakis:2008hj}
P.~Maragakis, M.~Spichty, and M.~Karplus,
\newblock J. Phys. Chem. B {\bf 112}, 6168 (2008).


\bibitem{Evans:1997np}
E.~Evans and K.~Ritchie,
\newblock Biophys. J. {\bf 72}, 1541 (1997).

\bibitem{hairpin} The DNA sequence is 5'-GCGAGCCATAATCTC
  ATCTGGAAACAGATGAGATTATGGCTCGC-3'. Pulling experiments were performed
  at ambient temperature ($24^{\circ}{\rm C}$) in a buffer containing
  Tris H-Cl pH 7.5, 1M EDTA and 1M NaCl. The DNA hairpin was hybridized
  to two dsDNA handles 29 base pairs long
  flanking the hairpin at both sides. Experiments
  were done in a dual-beam miniaturized optical tweezers with
  fiber-coupled diode-lasers  (845 nm wavelength) that produce a piezo controlled movable optical trap and measure force using
  conservation of light momentum.  The experimental setup is based on:
  C.\ Bustamante and S.\ B.\ Smith, Nov.\ 7, 2006. U.S.\ Patent
  7,133,132,B2.

\bibitem{Shirts:2003ok}
M.~R. Shirts, E.~Bair, G.~Hooker, and V.~S. Pande,
\newblock \prl {\bf 91}, 140601 (2003).

\bibitem{Bennett:1976ju}
C.~H. {Bennett},
\newblock J. Comp. Phys. {\bf 22}, 245 (1976).






\end{thebibliography}

\clearpage
\hspace*{-2.5cm} \includegraphics{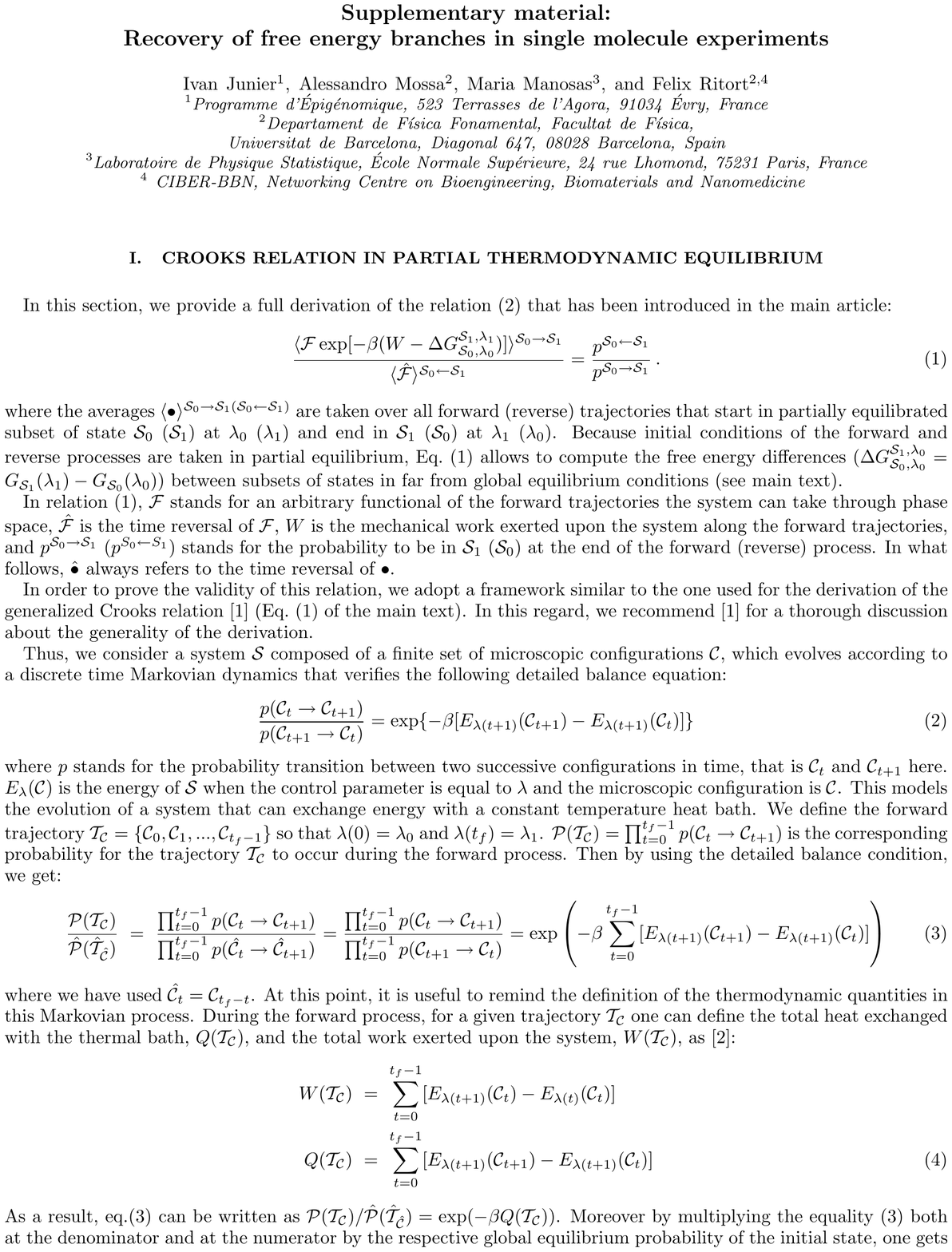}
\clearpage
\hspace*{-2.5cm} \includegraphics{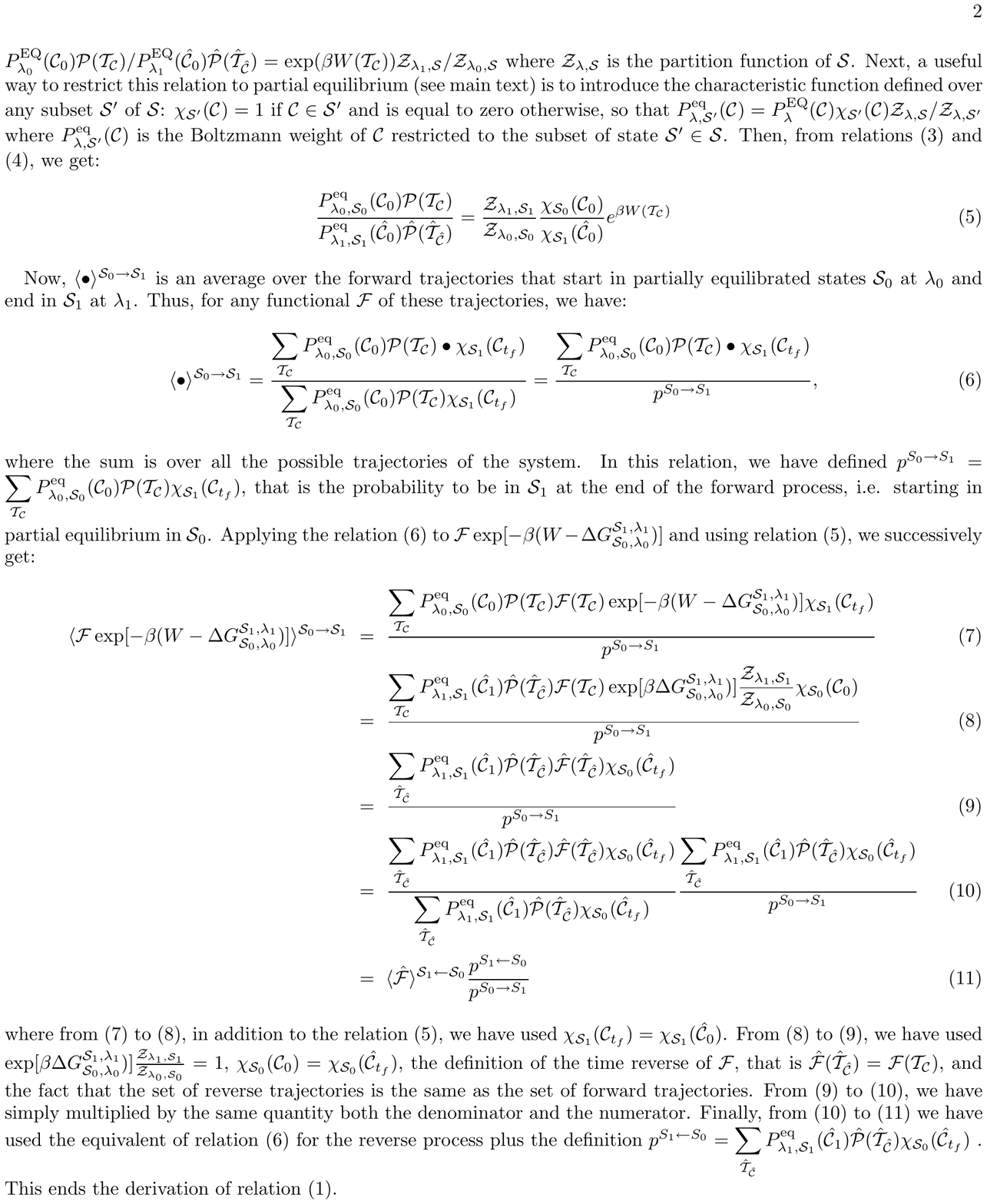}
\clearpage
\hspace*{-2.5cm} \includegraphics{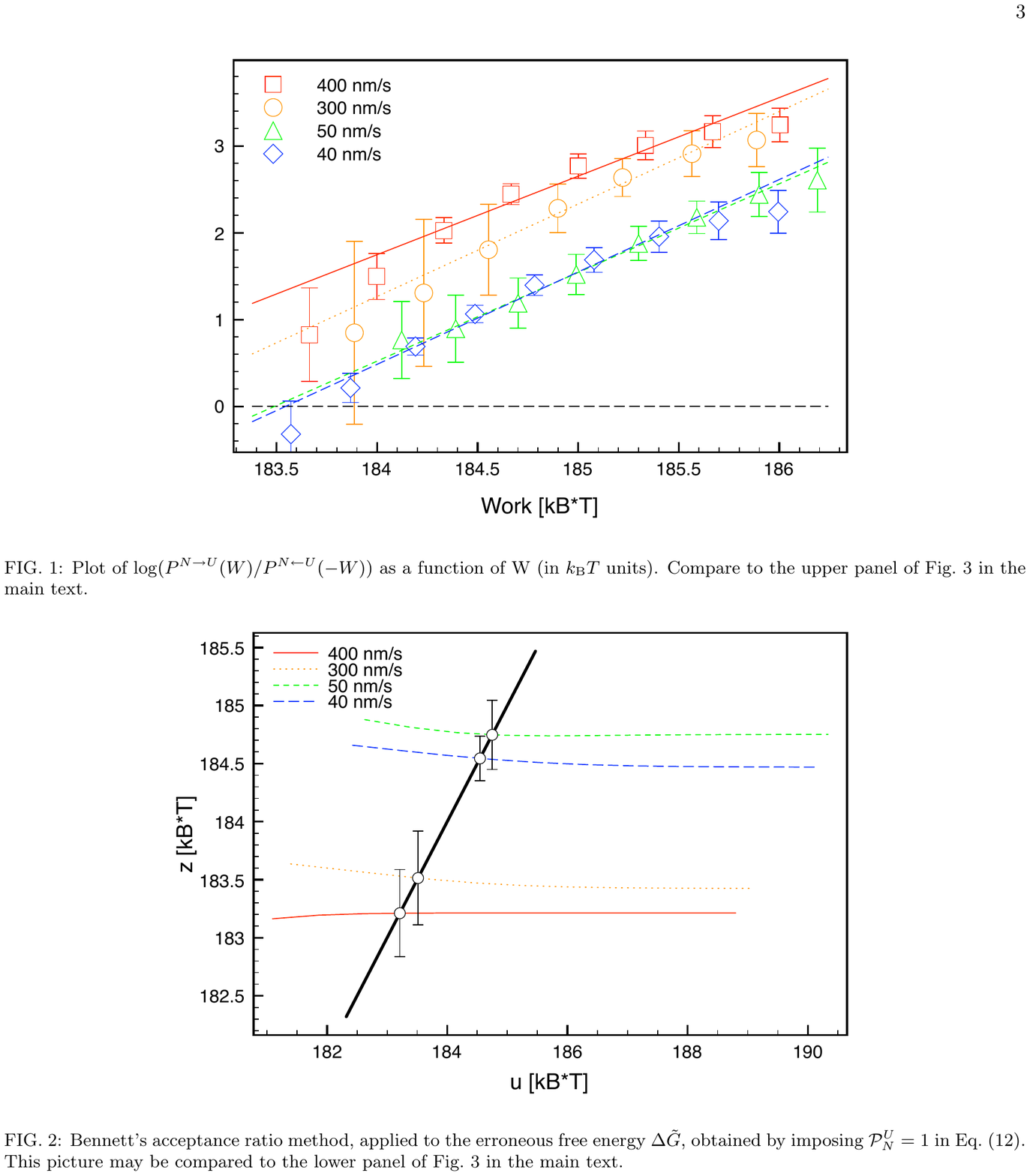}
\clearpage
\hspace*{-2.5cm} \includegraphics{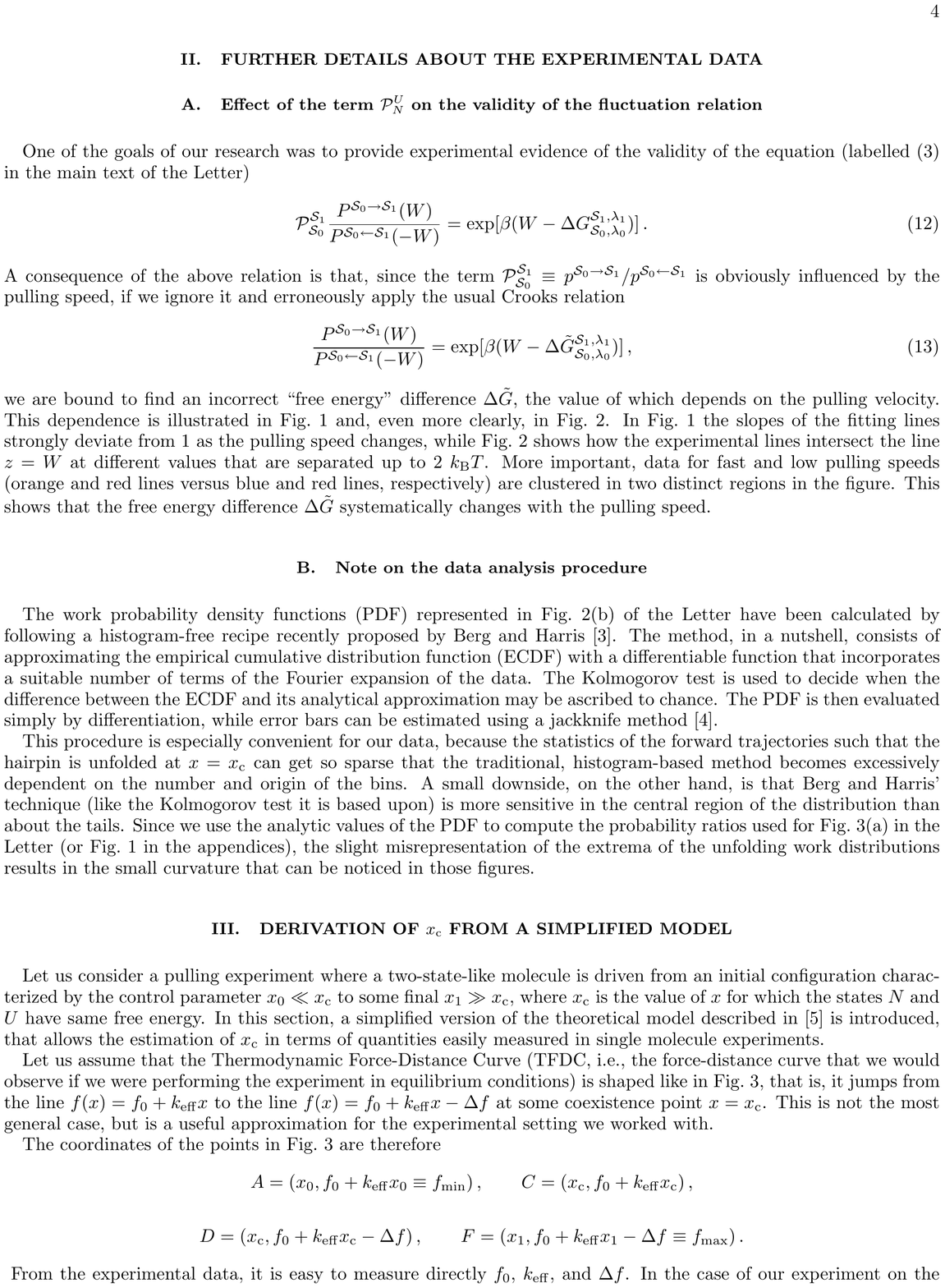}
\clearpage
\hspace*{-2.5cm} \includegraphics{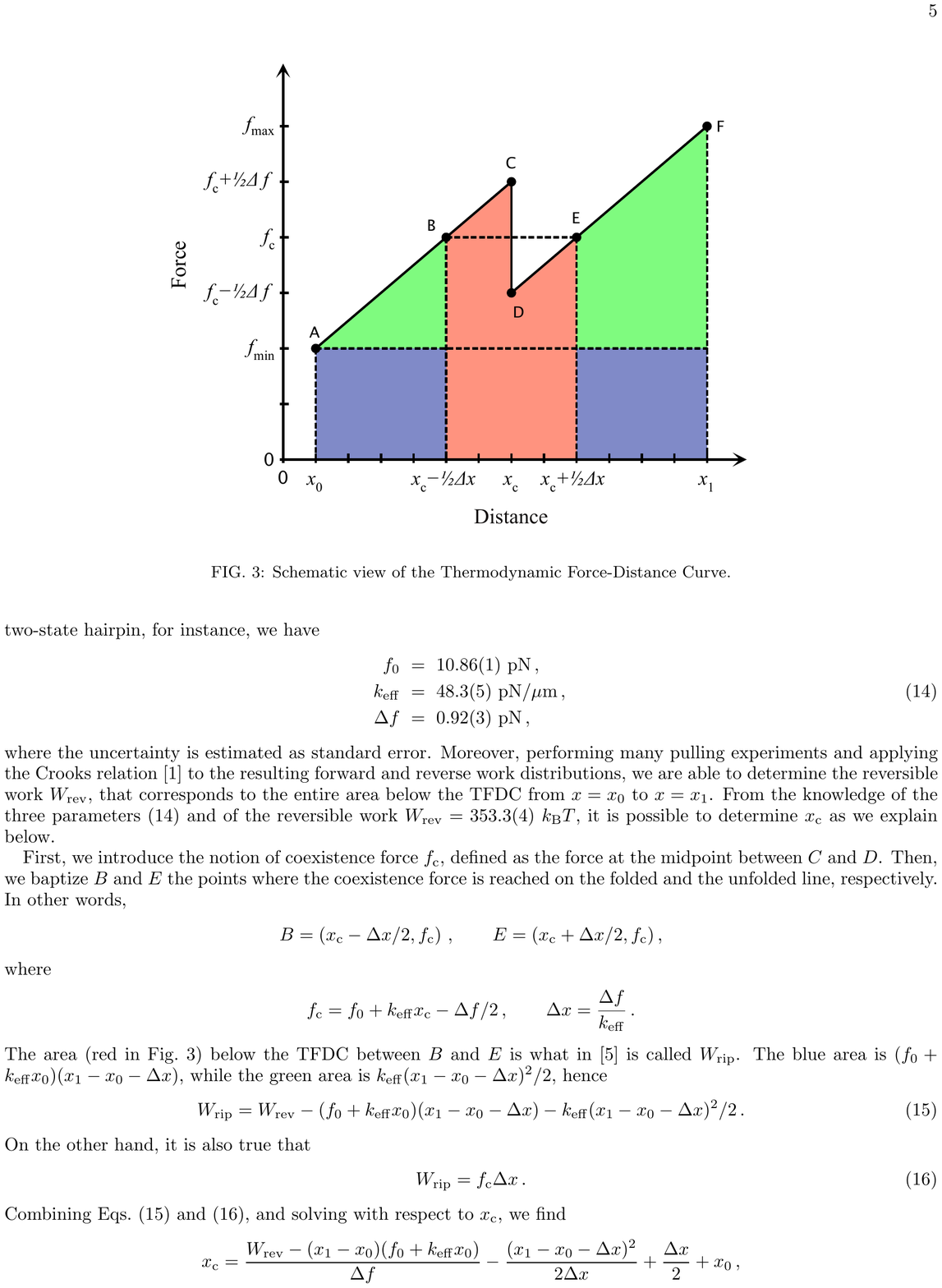}
\clearpage
\hspace*{-2.5cm} \includegraphics{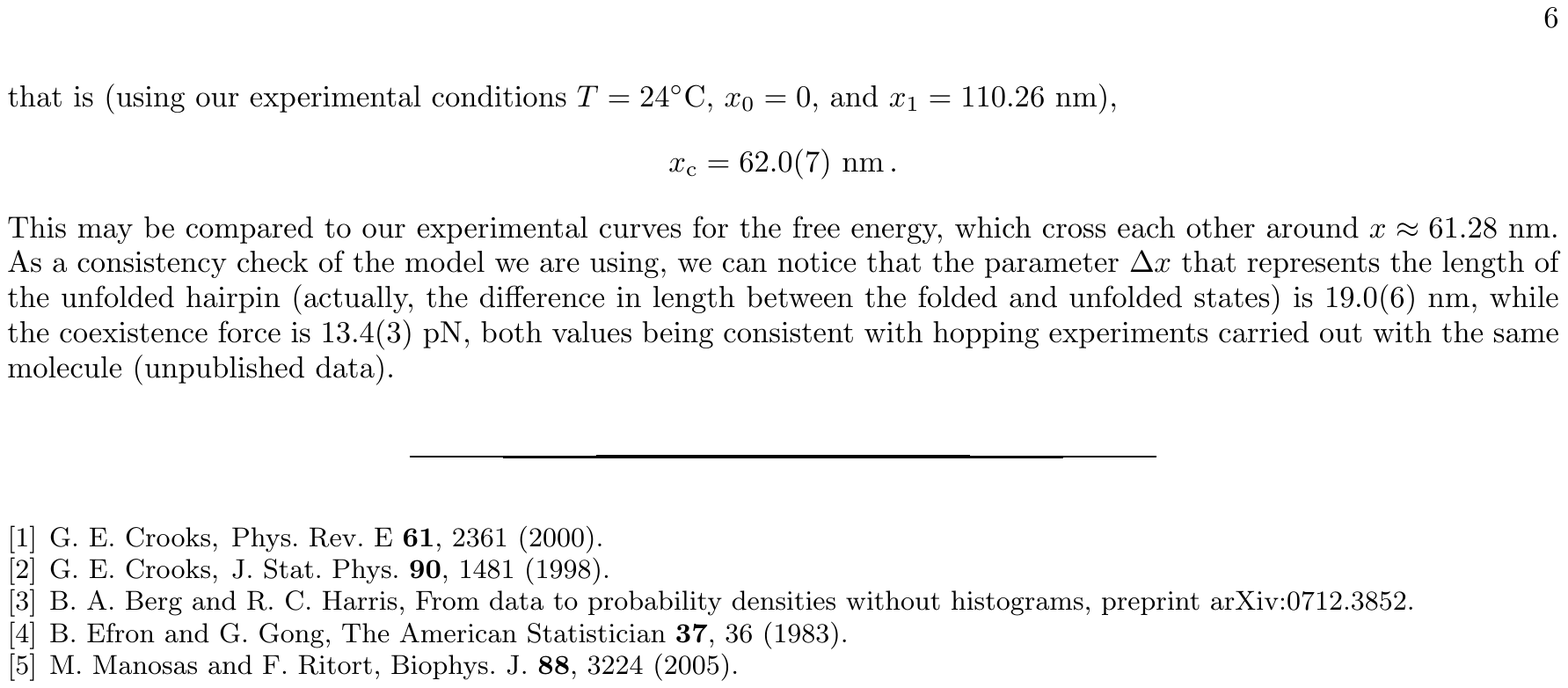}

\end{document}